\documentclass[conference]{IEEEtran}
\usepackage{subcaption}
\usepackage{caption}  

\usepackage{cite}
\usepackage{amsmath,amssymb,amsfonts}
\usepackage{algorithmic}
\usepackage{subcaption}
\usepackage{enumitem}
\usepackage{graphicx}
\usepackage{algorithm}                
\usepackage{algorithmic}
\usepackage{amsthm}
\usepackage{amsmath, amssymb, mathtools}
\usepackage{bm}
\newcommand{\vect}[1]{\bm{#1}}
\usepackage{booktabs}
\usepackage{graphicx}
\usepackage{textcomp}
\usepackage{xcolor}
\usepackage{lipsum}
\def\BibTeX{{\rm B\kern-.05em{\sc i\kern-.025em b}\kern-.08em
    T\kern-.1667em\lower.7ex\hbox{E}\kern-.125emX}}
\begin{document}

\title{Argus: Token Aware Distributed LLM Inference Optimization}

\author{
    \IEEEauthorblockN{Panlong Wu*}
    \IEEEauthorblockA{\textit{CUHK(SZ)}\\
    Shenzhen, China}
    \and
    \IEEEauthorblockN{Yifei Zhong*}
    \IEEEauthorblockA{\textit{CUHK(SZ)}\\
    Shenzhen, China}
    \and
    \IEEEauthorblockN{Danyang Chen}
    \IEEEauthorblockA{\textit{CUHK(SZ)}\\
    Shenzhen, China}
    \and
    \IEEEauthorblockN{Ting Wang}
    \IEEEauthorblockA{\textit{CUHK(SZ)}\\
    Shenzhen, China}
    \and
    \IEEEauthorblockN{Fangxin Wang}
    \IEEEauthorblockA{\textit{CUHK(SZ)}\\
    Shenzhen, China}
}

\maketitle

\begin{abstract}
Large Language Models (LLMs) are rapidly being integrated into real-world applications, yet their autoregressive architectures introduce significant inference time variability, especially when deployed across heterogeneous edge-cloud systems. Existing solutions largely neglect the dynamic, stochastic, and heterogeneous nature of such environments, often ignoring the impact of variable output token lengths and device diversity. In this work, we present Argus, the first token-aware distributed edge-cloud LLM inference framework that conducts efficient task offloading. Argus features a Length-Aware Semantics (LAS) module, which predicts output token lengths for incoming prompts using a fine-tuned language model with token-length-sensitive feature modulation, enabling precise estimation. Building on this, our Lyapunov-guided Offloading Optimization (LOO) module formulates long-term Quality-of-Experience optimization that explicitly considers both LLM prefilling and decoding costs. We introduce a novel Iterative Offloading Algorithm with Damping and Congestion Control (IODCC) to effectively solve the resulting integer nonlinear programming problem under time-varying constraints. Extensive theoretical and empirical evaluations demonstrate that Argus achieves robust performance and superior efficiency in highly dynamic, heterogeneous settings.
\end{abstract}


\let\thefootnote\relax\footnotetext{Authors with $*$ contributed equally to this work.}

\section{Introduction}
Recent advances in Large Language Models (LLMs) have positioned them as a transformative force within artificial intelligence, showcasing unprecedented capabilities in understanding and reasoning across various domains. Typically trained on trillions of tokens and comprising billions of parameters, these models have extensive knowledge that enables them to capture complex patterns and relationships effectively.

Due to their large number of parameters, LLMs require substantial computational resources and are typically deployed in the cloud. In real-world scenarios, different tasks have varying requirements for latency and accuracy. Centralizing LLMs in the cloud can significantly increase latency, which is detrimental to latency-sensitive tasks. A promising distributed deployment solution involves deploying small-scale LLMs at the edge and large-scale LLMs in the cloud, and coordinating them. Advances in technology have equipped many modern edge servers with GPUs (such as the NVIDIA Jetson Orin), enabling them to deploy small-scale LLMs as edge computing devices. Cloud clusters 

\begin{figure*}[t]
    \centering
    \begin{subfigure}[t]{0.48\textwidth}
        \centering
        \includegraphics[width=\textwidth]{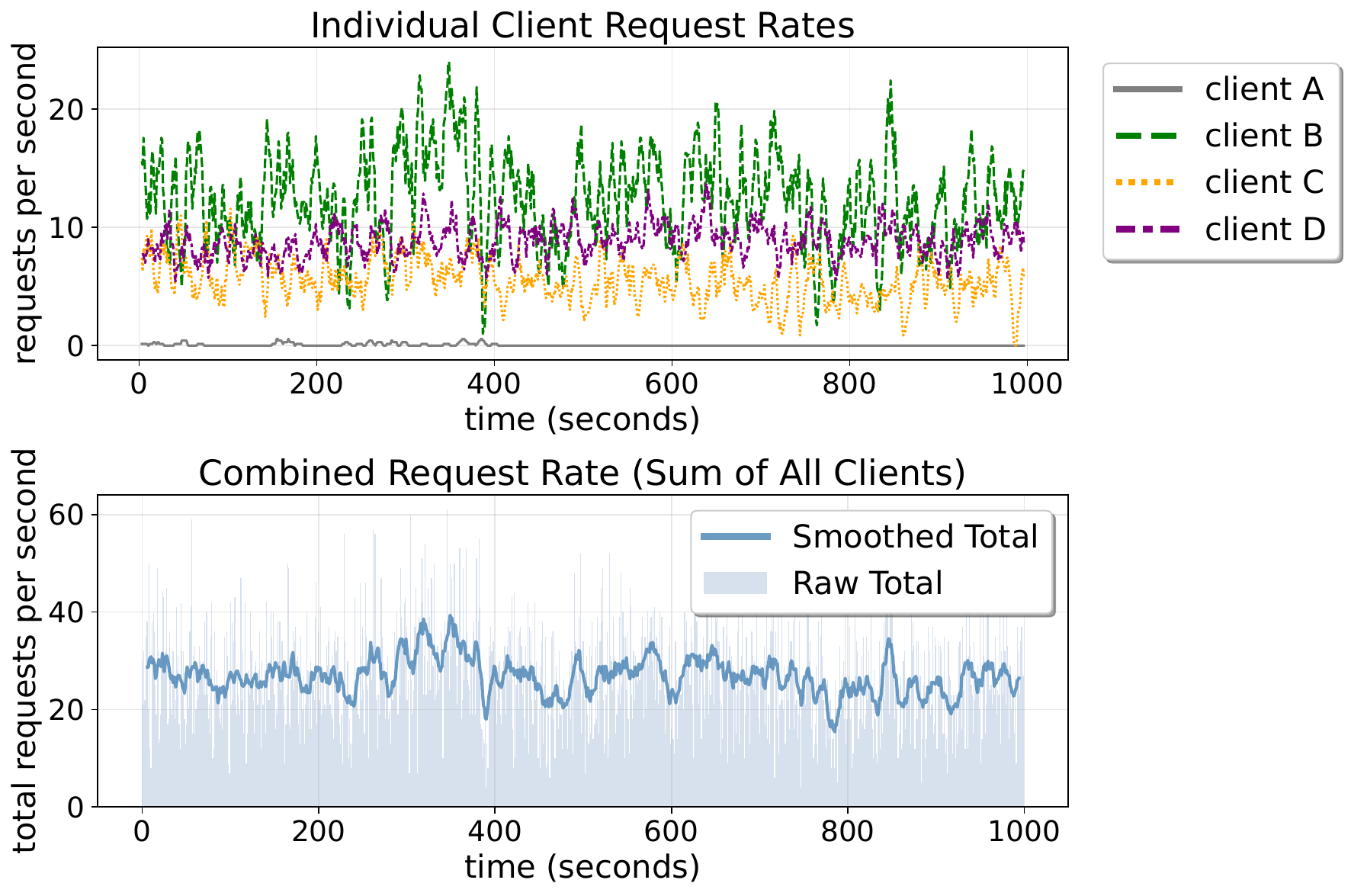}
        \caption{LLM requests data sample from different clients}
        \label{fig:requests_data}
    \end{subfigure}
    \hfill
    \begin{subfigure}[t]{0.48\textwidth}
        \centering
        \includegraphics[width=\textwidth]{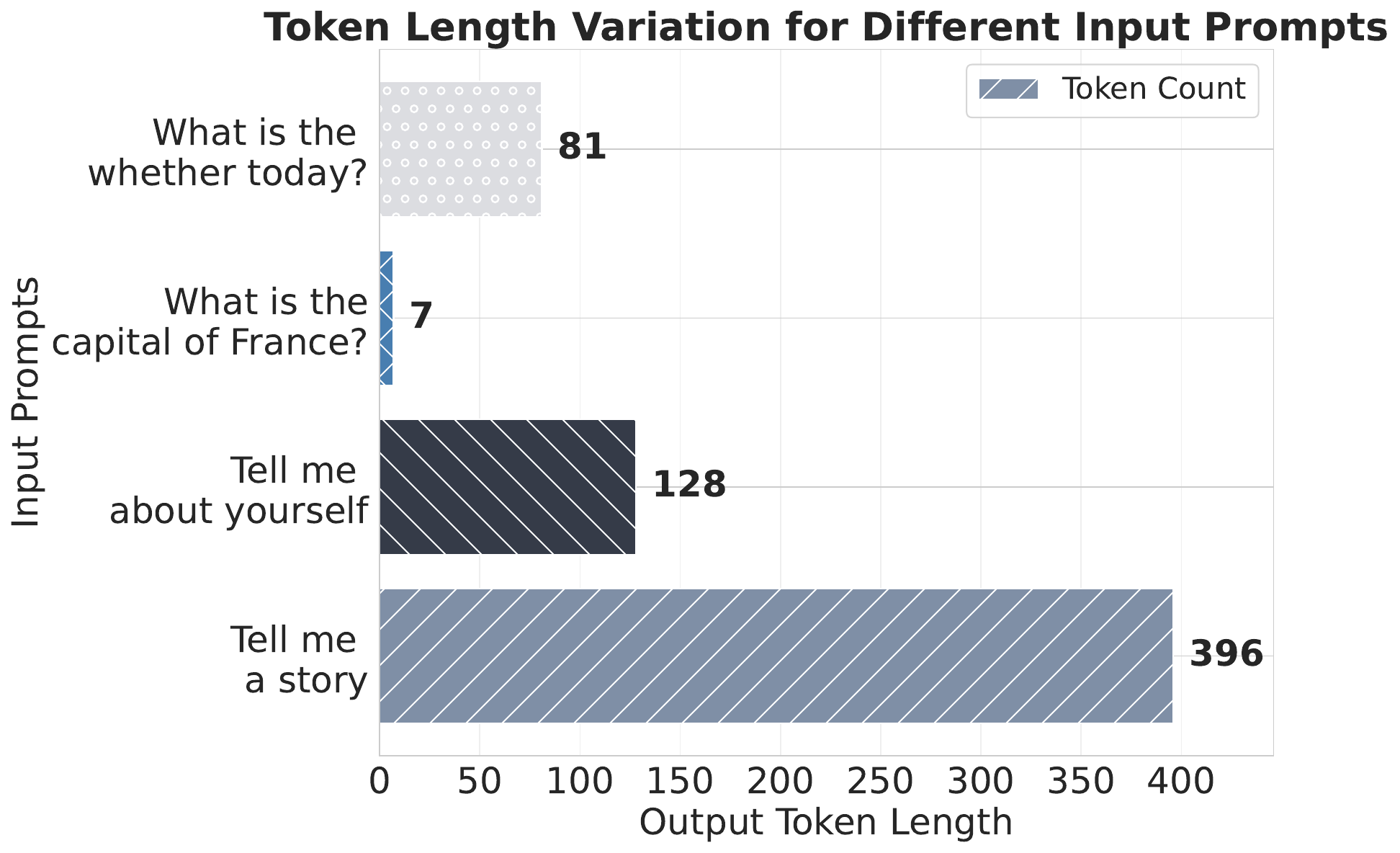}
        \caption{Different output token length on different queries for the same LLM}
        \label{fig:various_token}
    \end{subfigure}
    \caption{Comparison of different data and token lengths related to LLM queries.}
    \label{fig:compare}
\end{figure*}

However, the collaborative inference deployment for small-scale LLMs and large-scale LLMs in edge server-cloud server systems still faces many challenges. 

\textbf{Decoder-only Transformer architecture causes significant inference time variations}. Traditional small models are mostly based on non-autoregressive architectures (such as Convolutional Neural Network and encoder-only transformer architecture). The inference time of these models primarily depends on their model size and structure. Consequently, for the same model, the inference time required for different problems does not vary much. In contrast, popular LLMs, such as LLaMA \cite{touvron2023llama} and DeepSeek\cite{guo2025deepseek}, typically employ autoregressive architectures, where the output is generated as a sequence of tokens. As depicted in Fig.\ref{fig:various_token}, for the LLMs, output token length exhibits significant variation depending on the semantic complexity and specificity of the input prompt. Consequently, inference time is heavily influenced by the number of tokens produced, leading to substantial variation—sometimes by factors of several or even dozens.

\textbf{Dynamic and stochastic environment}. 
In real-world deployments, both the volume of incoming inference queries and the set of active clients can fluctuate dramatically over time, often compounded by highly variable and unstable network conditions. For example, real request data from various clients collected by Alibaba \cite{xiang2025servegen} (see Fig.~\ref{fig:requests_data}) demonstrates significant fluctuations in request rates, reflecting pronounced volatility in both individual and aggregate client activity. This is \textit{especially difficult combined with the highly variant token length of different clients}. Existing studies often optimize system performance at a static point in time, neglecting the stochastic and time-varying nature of real-world environments and failing to address long-term temporal variability and randomness.

\textbf{Heterogeneous Environment}.
Distributed systems typically comprise a diverse mix of heterogeneous edge and cloud servers, each with distinct computational capacities and communication characteristics. This diversity further complicates resource allocation and system management, as devices are subject to varying constraints. Many previous works overlook these heterogeneous, long-term average constraints across devices. Thus, there is an urgent need for optimization frameworks that jointly capture the challenges posed by temporal variability, environmental randomness, and device heterogeneity to ensure robust, efficient, and sustainable distributed LLM inference.

To address this gap, we introduce Argus, \textbf{the first token-aware distributed edge cloud inference optimization framework} for LLMs that can efficiently offload requests with varying characteristics to heterogeneous computing devices, leveraging the intrinsic autoregressive property of LLMs. Our design consists of two modules. The first is the Length-Aware Semantics (LAS) module, which adaptively predicts the expected output token length for each input prompt by leveraging a fine-tuned pre-trained language model with a novel feature modulation module tailored for token-length sensitivity. By accurately \textbf{estimating token lengths before generation}, the LAS module enables fine-grained request profiling, which is essential for further tasks offloading. The second is the Lyapunov-guided Offloading Optimization (LOO) module based on Lyapunov optimization\cite{neely2010stochastic}. We formulate the long-term Quality-of-Experience (QoE) optimization with specially consideration of the \textbf{prefilling and decoding cost of LLM} and transform it into a series of per-timeslot minimization problems. Inspired by control theory, we propose a novel Iterative Offloading Algorithm with Damping and Congestion Control (IODCC), which addresses the per-timeslot Integer Nonlinear Programming (INLP) problem through an iterative feedback loop. The algorithm incorporates a dynamic, congestion-aware cost function and a damped update mechanism to enhance stability and convergence.

Our contribution can be summarized as follows:
\begin{itemize}[topsep=0pt, partopsep=0pt, itemsep=0pt, parsep=0pt]

\item We present the first optimization framework that explicitly models the autoregressive token-generation behavior of LLMs in edge-cloud systems. 

\item We formulate the distributed LLM inference problem as a scholastic optimization problem  with consideration of prefilling and decoding cost of LLM. We propose a LOO framework that transforms the intractable long-term distributed LLM inference optimization problem into per-time slot optimization problems and further solves this INLP by a low complexity IODCC algorithm.

\item We provide detailed theoretical performance analysis of our algorithm and conduct extensive experiments on various conditions to further prove its effectiveness.
\end{itemize}

\begin{figure*}[thbp]
    \centering
    \includegraphics[width=0.8\textwidth, height=5cm, keepaspectratio]{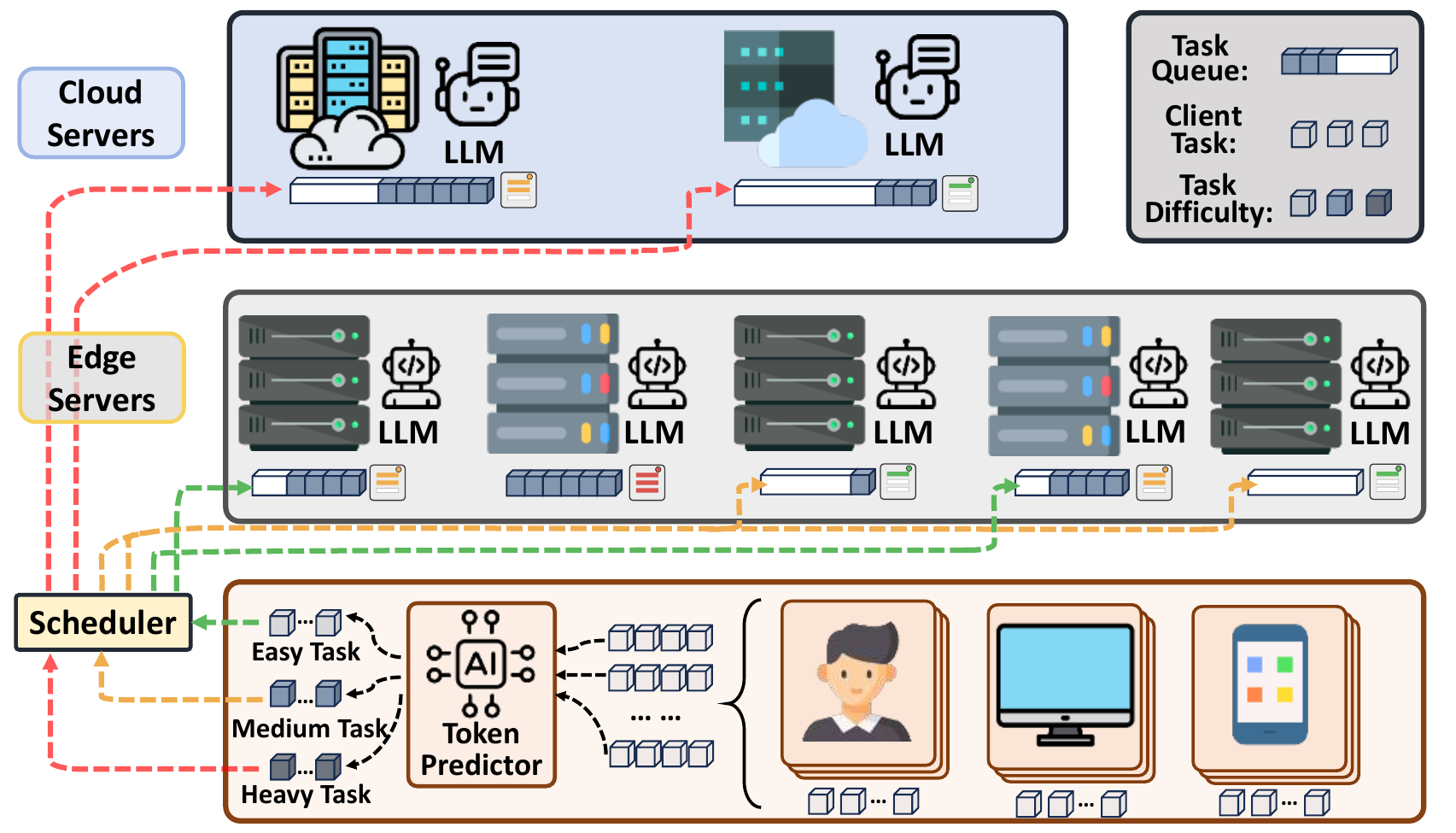}
    \caption{System overview of the proposed distributed LLM system}
    \label{fig:prediction}
\end{figure*}

\section{Related Work}
\subsection{Resource Allocation in Distributed Systems}
Many works have been done related to resource allocation in distributed systems. In \cite{wang2020machine}, the authors propose a learning-centric power allocation (LCPA) scheme for edge machine learning systems, which optimizes power allocation to maximize learning performance rather than communication throughput. The authors formulate the problem as a nonconvex nonsmooth optimization problem and solve it using a majorization minimization (MM) framework. In \cite{chen2018efficient}, the authors propose a comprehensive framework for efficient resource allocation in on-demand mobile-edge cloud computing, including a resource-efficient computation offloading mechanism for users and a joint communication and computation (JCC) resource allocation mechanism for network operators. 
In \cite{ren2018latency}, the authors investigate latency optimization for resource allocation in mobile-edge computation offloading (MECO) systems, focusing on minimizing end-to-end latency through joint communication and computation resource allocation. It explores three computation models—local compression, edge cloud compression, and partial compression offloading—deriving optimal resource allocation strategies and demonstrating significant latency reduction through numerical results. However, all these works is quite different from distributed LLM inference, which will have unique challenges because of their autoregressive property.

\subsection{LLM inference optimization}
Various works have been done related to LLM inference optimization. 
In \cite{xiao2023smoothquant}, the authors propose a novel quantization method that can significantly improve the performance when models are quantized into 8 bits. 
In \cite{kwon2023efficient}, the authors propose a revolutionary PagedAttention memory management mechanism. The key-value cache is paged just like how an operating system manages physical memory, allowing multiple sequences to efficiently share GPU memory, enabling dynamic allocation and reclamation. In\cite{liu2024cachegen}, the authors propose CacheGen, which is a novel system designed to accelerate LLM serving by compressing and streaming KV caches efficiently, significantly reducing context-loading delays and network bandwidth usage while maintaining high response quality. By leveraging custom tensor encoding and adaptive streaming techniques, CacheGen achieves substantial improvements in KV cache size reduction and time-to-first-token (TTFT) across various models and datasets, making it a promising solution for fast and efficient LLM inference. 
In \cite{yao2025cacheblend}, the authors propose CacheBlend, a novel system designed to optimize the inference speed of LLMs in retrieval-augmented generation (RAG) scenarios by efficiently fusing precomputed key-value caches of multiple text chunks.
In \cite{zheng2024sglang}, the authors propose SGLang, a system for efficient execution of complex language model programs, which includes a frontend language with primitives for generation and parallelism control, and a runtime with novel optimizations like RadixAttention for KV cache reuse and compressed finite state machines for faster structured output decoding. 
However, all these works do not consider the edge cloud distributed inference scenario, which introduces many complexities to be optimized.

\section{System Model}

We consider a scenario, where there are $K$ different types of tasks, which require LLMs deployed on the computation devices for further task completion. We aim to optimize the strategy over sequential time frames, spanning a total duration of $T$. At each time slot $t$, let $E_t$ denote the total number of queries, and $M_t$ denote the set of clients offloading their tasks. The system consists of $N$ edge servers nearby and $U$ cloud servers far away.

For each client $m\in M_t$ at time slot $t$, let $E_m^{(t)}$ be the set of tasks it owns at time $t$. The communication delay for client $m$ offloading task $e\in E_m^{(t)}$ to the computation device $j$ at time $t$ is given by:
\begin{equation}
    \kappa_{m,j}^e(t) = a_{m,j}^e(t) \left(\frac{\digamma_e(t)}{r_{m,j}(t)} +  \eta_{m,j}(t)\right)
\end{equation}
where \( a_{m,j}^e(t) \) is a binary variable indicating whether client $m$ at time $t$ offloads task $e$ to the computation device $j$, \( \digamma_e(t) \) is the data size of task $e$ at time $t$, \( r_{m,j}(t) \) is the communication rate between client $m$ and computation device $j$ at time $t$, and $\eta_{m,j}(t)$ denotes the network delay caused by data propagation and network protocol processes at time $t$.

The communication rate \( r_{m,j}(t) \) is determined by the link conditions. Specifically, the rate between client $m$ and edge server $j$ at time $t$ is denoted $r_{m,j}^{\text{CE}}(t)$, and for a cloud server it is $r_{m,j}^{\text{CC}}(t)$.

For a client $m$ to successfully connect to computation device $j$ at time $t$, the communication rate must exceed a minimum threshold \( r_{\min}(t) \):

\begin{equation}
    a_{m,j}^e(t) \leq \mathbb{I}(r_{m,j}(t) > r_{\min}.
\end{equation}

Inference in contemporary LLMs is commonly structured as a two-stage process: the prefilling phase, in which the entire input prompt is processed to initialize the key-value cache; and the decoding phase, where output tokens are generated sequentially by leveraging cached representations for efficient autoregressive generation.
We assume that each query has a distinct workload, denoted by $q_e(t)$, representing the prefilling and decoding stages duration of task $e$ at time $t$. The offloading decision is represented by a binary variable \( a_{m,j}^e(t) \in \{0,1\} \), indicating whether the task $e$ from the \( m \)-th client is offloaded to the \( j \)-th computation device at time frame \( t \).

At each time slot, each client random has several tasks arrive or not. We assume that each query can be offloaded only to one computation device:

\begin{equation}
    \sum_{j=1}^{N+U} a^e_{m,j}(t) = 1, \quad \forall e\in E_m^{(t)}, t\in[1,T], m\in[1,M].
\end{equation}

We assume there is computational constraints that the long-term time-averaged total computation time of tasks allocated to device $j$ should not exceed its computation capacity:
\begin{equation}
\lim_{T\to\infty} \frac{1}{T} \sum_{t=1}^{T} \sum_{m\in M_t} \sum_{e\in E_m^{(t)}} a^e_{m,j}(t) \frac{q_e(t)}{f_j} \leq \Upsilon_j, \quad \forall j
\end{equation}
where \( \Upsilon [j]\) denotes the computation threshold of device $j$, and \( f_j \) denotes the computational capacity of computation device $j$.

We construct a task queue on each devices which follows the first in first out strategy. The computation delay of a task consists of the processing time of the tasks already in the queue of the computation device, tasks offloaded in the same time slot and arrive earlier than this task, and the computation time of itself. Therefore the computaion delay of task $e$ offloaded to computation device $j$ at time $t$ can be represented as 

\begin{equation}
\tau_{e,j}(t) = \frac{1}{f_j} \left[ Q_j(t) + \sum_{\substack{e' \in O_j(t) \\ e' \prec e}} q_{e'}(t) + q_e(t) \right]
\label{latency}
\end{equation}
where $O_j(t)$ denotes the the set of all new tasks that are offloaded to device $j$ in the current time slot $t$. $c_e$ denotes the computation workload of task $e$.
$e'$ denotes tasks that offloaded in the same time
slot and arrive earlier than task $e$.

As different tasks have different requirement of accuracy and latency, we denote $\alpha_k$ to be the delay sensitivity of task $k$ and $\beta_k$ to be the accuracy sensitivity of task $k$. $\delta$ denotes the weight factor of the delay and accuracy. $\phi_{m,e,j}(t)$ denotes the accuracy of task $e$ offloaded from client $m$ to computation device $j$.

The optimization problem can be formulated as
\begin{subequations}
\begin{align}
\min_{\mathbf{a}(t)} \quad & \lim_{T \to \infty} \frac{1}{T} \sum_{t=1}^{T} \Biggl\{ \sum_{m=1}^{M} \sum_{e \in E_m} \sum_{j=1}^{N+U} a^e_{m,j}(t) \cdot \nonumber \\
& \qquad \qquad \left[ \alpha_e \tau_{m,e,j}(t) - \delta \beta_e \phi_{m,e,j}(t) \right] \Biggr\}\\
\text{s.t.} \quad & \sum_{j=1}^{N+U} a^e_{m,j}(t) = 1,  \forall m \in [1,M], \forall e \!\in \! E_m, \forall t \!\in\! [1,T] \label{1} \\
& \lim_{T \to \infty}\frac{1}{T}\sum_{t=1}^{T}\frac{\sum_{e=1}^E a^e_{j}(t) \cdot q_e}{f_j} - \!\Upsilon[j] < 0, \\ \nonumber
&\qquad\forall j\in[1,N\!+\!U]  \\
& a^e_{m,j}(t) \in {0, 1}, \forall m, e \in E_m, j, \forall t \in [1,T] \label{3} \\
& a^e_{m,j}(t) \leq \mathbb{I}(r_{m,j}^{CE}(t) > r_{\min}), \label{5} \\  \nonumber
&\qquad  \forall m, e \in E_m, j \in [1,N], \forall t \in [1,T]  \\
& a^e_{m,j}(t) \leq \mathbb{I}(r_{m,j}^{CC}(t) \!> \!r_{\min}), \label{6} \\ \nonumber
&\qquad \!\forall m, e \in E_m, j \in [N\!+\!1\!, \!N\!+\!U], \forall t \in [1,T]. \\ \nonumber
\end{align}
\end{subequations}

\subsection{token length prediction}

The architectural and operational paradigms of LLMs fundamentally distinguish them from conventional deep learning models like CNN-based and MLP-based architectures. While traditional models exhibit static computational complexity determined by fixed network sizes and architectures, LLMs based on autoregressive Transformer architectures introduce dual-dimensional workload dynamics. The inference time of these models scales not only with the parameter count and model architecture but critically depends on the generated token sequence length, a unique characteristic arising from their autoregressive token generation process. However, previous works of distributed deep neural network inference\cite{zhang2025breaking}\cite{ghosh2023energy} do not consist of this important feature in distributed LLM inference optimization which leads to suboptimalness.  

\begin{figure}[htbp]
    \centering
    \includegraphics[width=2.4\textwidth, height=5cm, keepaspectratio]{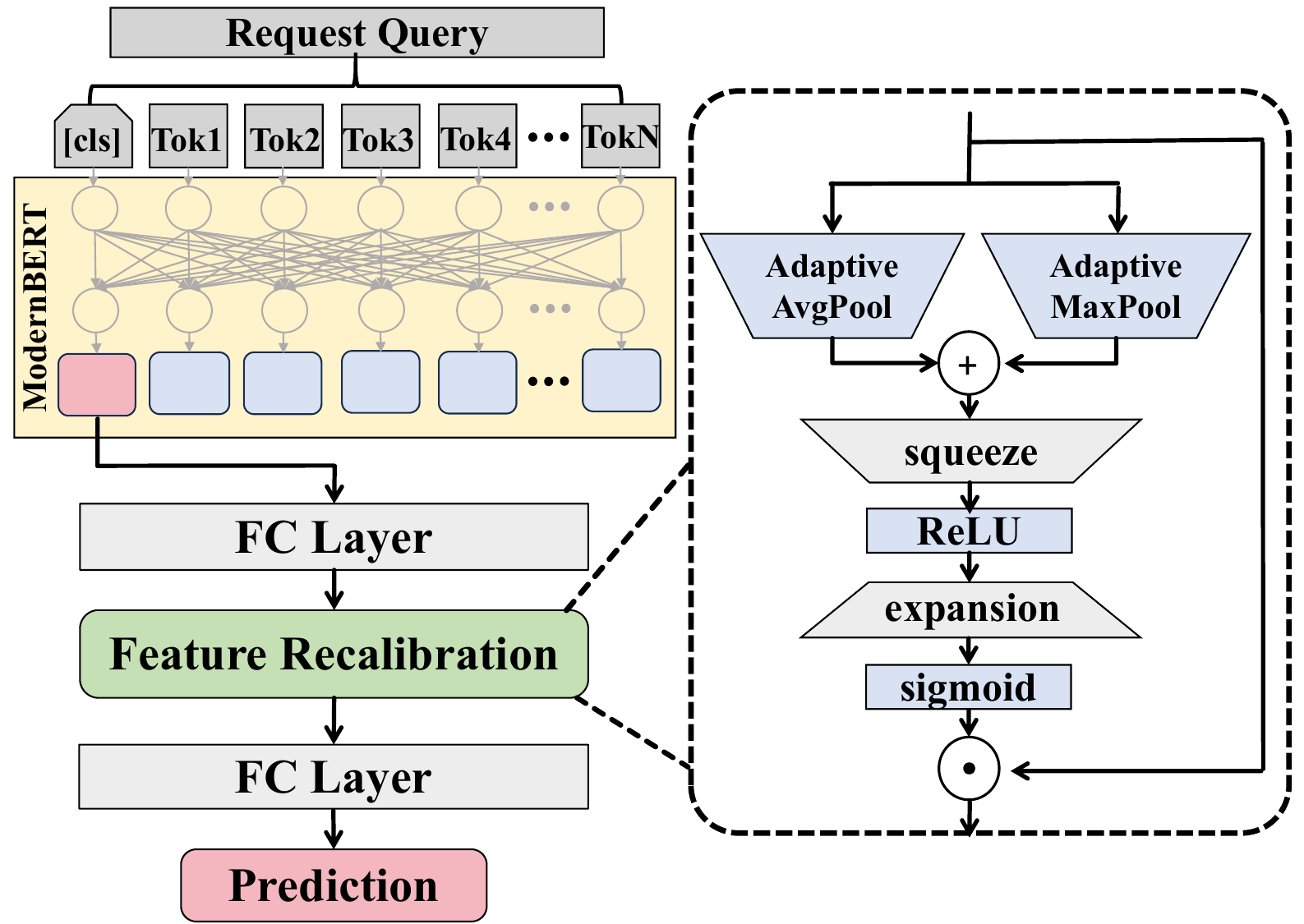}
    \caption{Design of LAS}
    \label{fig:prediction}
\end{figure}

To fill this gap, we proposed an LAS algorithm to obtain a precise token prediction. The design of our algorithm is shown in Fig. \ref{fig:prediction}. To accurately predict the output token length of LLMs for various prompts, we propose a token prediction enhancement framework built upon a pre-trained language model. The core of LAS is to map a user's input prompt to a scalar value representing the predicted output length. Because of the fact that training a model from scratch to comprehend user prompts and predict response length would necessitate vast amounts of labeled data and immense computational resources. More critically, such a model would struggle to capture the complex semantic and inferential nuances of natural language, which are crucial for determining the expected response length (e.g., "explain in detail" vs. "list briefly"). Therefore, our approach is to utilize the powerful pretrained language model modern-bert\cite{warner2024smarter} to do further fine-tuning. Through extensive training on massive text corpora, it has acquired a deep understanding of linguistic structures and world knowledge, which benefits the prediction of the output token length. 

To enhance the pretrained language model to have the ability to do token length prediction, we design a LAS module that insert to the pretrained model and tunes these modules. The philosophy of LAS is that Not all features within the semantic vector $\boldsymbol{z}$ are equally relevant for predicting output length. For instance, features related to the topic might be less important than features indicating an instruction to "be concise" or "explain in detail." To address this, we introduce a Length-Aware Semantic (LAS) module, which is inspired by Squeeze-and-Excitation (SE) networks \cite{hu2018squeeze} commonly used in computer vision, but tailored for selectively modulating features that are most indicative of the desired output length in natural language prompts. This module dynamically re-weights feature importances to enhance the model’s sensitivity to prompt cues relevant to token count prediction. The FR module operates in three steps:

1. Squeeze: This step aggregates global information from the feature vector. We employ both average-pooling and max-pooling to capture both the global statistics and the most salient feature signals. Given the feature vector $\boldsymbol{z}$, the pooled descriptor $\boldsymbol{s}$ is computed as:
$$
\boldsymbol{s} = \text{AdaptiveAvgPool}(\boldsymbol{z}) + \text{AdaptiveMaxPool}(\boldsymbol{z}).
$$
This combination provides a richer and more robust summary of the features.

2. Excitation: The squeezed descriptor $\boldsymbol{s}$ is then fed through a small bottleneck structure composed of two fully-connected (FC) layers. This learns a non-linear, cross-feature interaction to generate a vector of attention-like weights.
$$
\boldsymbol{e} = \sigma(\boldsymbol{W}_{exp}(\text{ReLU}(\boldsymbol{W}_{sq}\boldsymbol{s}))).
$$
Here, $\boldsymbol{W}_{sq} \in \mathbb{R}^{d_{squeeze} \times d_{hidden}}$ and $\boldsymbol{W}_{exp} \in \mathbb{R}^{d_{hidden} \times d_{squeeze}}$ are the weights of the two FC layers, with $d_{squeeze} \ll d_{hidden}$ forming the bottleneck. The sigmoid function $\sigma(\cdot)$ scales the resulting weights to the range [0, 1], making them suitable for attentional modulation.

3. Recalibrate: The final step involves applying the learned excitation weights $\boldsymbol{e}$ to the original feature vector $\boldsymbol{z}$ through element-wise multiplication,
$$
\boldsymbol{z'} = \boldsymbol{z} \odot \boldsymbol{e}.
$$
This operation selectively amplifies important features and diminishes irrelevant ones, producing a recalibrated feature vector $\boldsymbol{z'}$ that is more attuned to the nuances of the length prediction task.

\begin{algorithm}[t]
\caption{Iterative Offloading Algorithm with Damping and Congestion Control (IODCC)}
\label{alg:iodcc}
\begin{algorithmic}[1]
    \STATE At the beginning of time slot $t$, observe the set of active tasks $\mathcal{M}_a(t)$, server computation capacities $\{f_j\}$, task properties $\{k_i, c_i\}$, and the virtual queue backlogs $\{Q_j(t)\}$.

    \STATE Determine the offloading decision $\mathbf{a}^*(t)$ by iteratively solving a sequence of assignment problems for $k=1, \dots, K_{\text{max}}$ or until convergence. In each iteration $k$:
    \begin{itemize}
        \item[\_] Construct a cost matrix $C^{(k)}$, where each element $C^{(k)}_{ij}$ is the estimated cost of assigning task $i$ to server $j$, defined as:
        $$ C^{(k)}_{ij} = \underbrace{C^{\text{base}}_{ij}}_{\text{Base Cost}} + \underbrace{P(\bar{L}^{(k-1)})}_{\text{Congestion Penalty}} $$
        where $C^{\text{base}}_{ij}$ includes transmission/computation delays and Lyapunov terms, and $P$ is a weight factor of the perceived server load $\bar{L}^{(k-1)}$ from the previous iteration.
        
        \item[\_] Obtain the current assignment $\mathbf{a}^{(k)}$ by solving the many-to-one assignment problem:
        $$ \min_{\mathbf{a}^{(k)}} \quad \sum_{i \in \mathcal{M}_a(t)} \sum_{j \in \mathcal{J}} C^{(k)}_{ij} a_{ij}^{(k)} \quad \text{s.t.} \quad \sum_{j \in \mathcal{J}} a_{ij}^{(k)} = 1, \forall i $$
        
        \item[\_] Update the perceived server load for the next iteration using a damping factor $\lambda_{\text{damp}}$:
        $$ \bar{L}_j^{(k)} \leftarrow (1 - \lambda_{\text{damp}}) \bar{L}_j^{(k-1)} + \lambda_{\text{damp}} \sum_{i \in \mathcal{M}_a(t)} a^{(k)}_{ij} C_i^{(k)} $$
    \end{itemize}

    \STATE Offload tasks according to the final converged decision $\mathbf{a}^*(t)$.

    \STATE Update the virtual queues from $Q_j(t)$ to $Q_j(t+1)$ based on the offloading decision $\mathbf{a}^*(t)$.

\end{algorithmic}
\end{algorithm}

\subsection{Algorithm Design}
In this paper, we aim to optimize the long-term average Quality of Experience (QoE) by utilizing reinforcement learning. Moreover, to satisfy the long-term constraint, we utilize the Lyapunov Optimization\cite{neely2010stochastic}, which introduce a lyapunov function to quantify the system's stability with respect to these constraints.

We denote 
\begin{equation}
y_j(t) =\!\!\!\! \sum_{m \in M_t} \!\sum_{e \in E^{(t)}_m}\! a^{e}_{m,j}(t) \frac{q_e(t)}{f_j} - \Upsilon_j, 
 \forall j \in \{1, \ldots, \!N \!+\! U\}.
\end{equation}\label{y_define}

\vspace{-1em}

For every $y_{j}(t)$, we define a corresponding $Q_{j}(t)$, which has the initial value of $0$.
\begin{equation}
Q_j(t+1) = max\{Q_j(t) + y_j(t), 0\}, j \in \{ 1,N+U\}.
\end{equation}

Then we can infer that 
\begin{equation}
    y_j(t) \leq Q_j(t+1) - Q_j(t) \label{equation}
\end{equation}
and we sum over all the time frame and get
\begin{equation}
    \sum_{t=0}^{T-1} y_j(t) \leq Q_j(T) - Q_j(0) = Q_j(T),  j \in [1, N+U],
\end{equation}
then we take expectations and get 
\begin{equation}
    \frac{1}{T} \sum_{t=0}^{T-1} \mathbb{E}[y_j(t)] \leq \frac{\mathbb{E}[Q_j(T)]}{T}, \quad j \in [1, N+U].
\label{eq11}
\end{equation}

To satisfy constraints (\ref{eq11}) we can set 
\begin{equation}
    \lim_{T \to \infty} \frac{\mathbb{E}[Q_j(T)]}{T} = 0,  j \in \{1, \dots, N+U\}.
\end{equation}

To solve this problem we define the Lyapunov function as 
\begin{equation}
    L(\Theta(t)) \triangleq \frac{1}{2} \sum_{j=1}^{N+U} Q_j(t)^2.
\end{equation}

The Lyapunov drift is defined as
\begin{equation}
    \Delta(\Theta(t)) \triangleq \mathbb{E}[L(\Theta(t+1))-L(\Theta(t)) \mid \Theta(t)].
\end{equation}

From \ref{equation}, we know that 
\begin{equation}
Q_j(t+1)^2 \leq (Q_j(t) + y_j(t))^2, \quad j \in \{1, N+U\}.    
\end{equation}

Based on this, summing over all $ K $ queues, we have
\begin{align}
    \frac{1}{2} \sum_{j=1}^{N+U} Q_j(t+1)^2 
    &\leq \frac{1}{2} \sum_{j=1}^{N+U} Q_j(t)^2 \notag \\
    &\quad + \frac{1}{2} \sum_{j=1}^{N+U} y_j(t)^2 + \sum_{j=1}^{N+U} Q_j(t)y_j(t)
\end{align}
\vspace{-1em}
and thus
\vspace{0.5em}
\begin{align*}
\Delta(\Theta(t)) &= \frac{1}{2} \sum_{j=1}^{N+U} Q_j(t+1)^2 - \frac{1}{2} \sum_{j=1}^{N+U} Q_j(t)^2 \\
&\leq \frac{1}{2} \sum_{j=1}^{N+U} y_j(t)^2 + \sum_{j=1}^{N+U} Q_j(t)y_j(t) .
\end{align*}
We define a $U\!B$ which is the upper bound of $\frac{1}{2} \sum_{j=1}^{N+U}y_j(t)^2$ and we can get 

\begin{equation}
   \Delta(\Theta(t)) \leq U\!B + \sum_{j=1}^{N+U} Q_j(t)y_j(t).
\end{equation}

We define 

\begin{equation}
    \zeta(t) = \Biggl\{ \sum_{m=1}^{M} \!\sum_{e \in E_m} \!\!\sum_{j=1}^{N+U}\! a^e_{m,j}(t) \cdot 
\left[ \alpha_e \tau_{m,e,j}(t) \!-\! \delta \beta_e \phi_{m,e,j}(t) \right] \Biggr\}.
\end{equation}

The optimization problem can be transformed into 

\begin{align}
\mathcal{P}1: \quad\min\limits_{\mathcal{A}_t}\operatorname*{lim}_{T\to\infty}&&& \!\!\frac{1}{T}\sum\limits_{t=1}^{T}\biggl[ \biggl( V \cdot\zeta(t) + \Delta(\Theta(t)) |\Delta(\Theta(t))\biggr] \nonumber \\
&&& s.t.\: \eqref{1},\; \eqref{3},\; \eqref{5},\; \eqref{6} .
\label{zeta_define}
\end{align}

We know that 
\begin{align}
V \cdot \zeta(t) + \Delta\big(\Theta(t)\big)
&\leq B + V \cdot \zeta(t) + \sum_{j=1}^{N+U} Q_j(t) y_j(t).
\label{eq:trans}
\end{align}

Then the problem can be transformed into
\begin{align}
    \mathcal{P}2:\min &\biggr[B + V \zeta(t)  + \sum_{j=1}^{N+U} Q_j(t)y_j(t) \biggr  |\Delta(\Theta(t))\biggr]\nonumber\\
     &s.t.\: \eqref{1},\; \eqref{3},\; \eqref{5},\; \eqref{6} .
\end{align}

According to \eqref{latency}, the cost of assigning task $e$ to device $j$ is affected by whether other tasks $e^{\prime}$ are also assigned to device $j$, the optimization problem of each time slot is an integer Nonlinear Programming (INLP) problem, which is NP-Hard and cannot be directly solved by commercial solvers. To solve this problem, we propose a novel Iterative Offloading Algorithm with Damping and Congestion Control (IODCC) algorithm. 

The core design philosophy of IODCC is to decompose the intractable INLP into a sequence of efficiently solvable Integer Linear Programs (ILPs). Instead of attempting to find the global optimum in a single, complex step, IODCC emulates a dynamic, multi-round process. In this process, the cost of utilizing a server increases in proportion to its congestion. This iterative feedback mechanism steers the assignments away from myopic, localized optima towards a globally efficient, load-balanced solution that effectively resolves resource contention among tasks. The algorithm is built upon three key pillars: a dynamic cost formulation with a congestion-aware penalty, an optimal assignment sub-solver, and a damped state update mechanism for stability. The complete procedure is formally outlined in Algorithm \ref{alg:iodcc}.

\textbf{1) Iterative Cost Formulation:} The central element of IODCC is the iterative construction of a cost matrix $C^{(k)}$ at each iteration $k$. Each cost element $C^{(k)}_{ij}$ is dynamically updated and comprises two main components. The first is a static Base Cost, which encapsulates the immediate costs of an assignment, such as communication/computation delay and the urgency dictated by the Lyapunov virtual queue backlog $Q_j(t)$. The second, and more crucial, component is a dynamic Congestion Penalty. This penalty quantifies the negative externality of intra-slot queuing as a function of the perceived server load, $\bar{L}^{(k-1)}$, from the previous iteration. By making the cost of assigning to a server dependent on its perceived congestion, the algorithm creates a powerful feedback loop that aggressively discourages overloading and promotes assignment diversity.

\textbf{2) Assignment via Integer Linear Programming:} 
With the cost matrix $C^{(k)}$ established for iteration $k$, the assignment subproblem becomes an ILP that can be solved to optimality efficiently using standard solvers. The ILP finds the assignment $a^{(k)}$ that minimizes the total cost $\sum_{i,j} C_{ij}^{(k)} a_{ij}^{(k)}$ subject to the constraint that each active task is assigned to exactly one server.

\textbf{3) Damped State Updates for Stability:} 
A naive iterative process where the perceived load is updated too aggressively is prone to oscillations, where tasks collectively shift between servers in successive iterations. To ensure stable and rapid convergence, IODCC incorporates a damped update mechanism. The perceived load for the next iteration, $\bar{L}_j^{(k)}$, is updated as a convex combination of the previous state and the new instantaneous load calculated from the current assignment $a^{(k)}$:

\begin{equation}
    \bar{L}_j^{(k)} \leftarrow (1-\lambda_{\text{damp}})\bar{L}_j^{(k-1)} + \lambda_{\text{damp}} \sum_{i \in \mathcal{M}_a(t)} a_{ij}^{(k)} c_i.
\end{equation}

The damping factor $\lambda_{\text{damp}} \in (0, 1]$ controls the smoothness of this update. It prevents the algorithm from overreacting to its own decisions, ensuring that the perceived server loads and the resulting assignment solution stabilize gracefully. This iterative process continues until the assignment matrix converges or a maximum number of iterations, $K_{\max}$, is reached. The final converged assignment $\mathbf{a}^*(t)$ is then used for the actual task offloading in time slot $t$.

\newtheorem{theorem}{Theorem}

\section{Theoretical Analysis}

In this section, we provide a rigorous theoretical analysis of the proposed algorithm, leveraging the foundational principles of Lyapunov optimization theory from \cite{neely2010stochastic}. Our objective is twofold: first, to establish a bound on the long-term time-averaged system objective, demonstrating its proximity to the optimal performance; and second, to prove that the proposed policy guarantees the satisfaction of the long-term computational constraints for all edge and cloud devices.

Let $\vect{\omega}(t)$ represent the collection of random events at timeslot $t$, including the set of arriving tasks $E_m^{(t)}$ for each client and the communication link conditions $r_{m,j}(t)$. We consider a stationary randomized policy $\vect{a}^*(\vect{\omega}(t))$ that depends only on the current random event $\vect{\omega}(t)$ and not on the virtual queue backlog $\vect{\Theta}(t)$. This policy is assumed to be optimal among all such stationary policies, achieving the following performance, for some optimal value $\zeta^*$:
\begin{align}
\mathbb{E}[\zeta(\vect{\omega}(t), \vect{a}^*(t))] &= \zeta^*, \\
\mathbb{E}[y_j(\vect{\omega}(t), \vect{a}^*(t))] &\leq 0, \quad \forall j \in \{1, \dots, N+U\}
\end{align}
where the expectation is taken over the distribution of $\vect{\omega}(t)$.

\begin{table*}[t]
\centering
\begin{minipage}{0.48\textwidth}
    \caption{\small Performance under different number of cloud servers}
    \label{tab:raw_rewards_u_configs}
    \centering
    \begin{tabular}{@{}lrrr@{}}
        \toprule
        \textbf{Algorithm}  & \textbf{U=15} & \textbf{U=20} \\ \midrule
        Ours  & \textbf{36602} & \textbf{30757} \\
        Baseline1 (Greedy-Accuracy) & -265297 & -213876 \\
        Baseline2 (Greedy-Compute)  & -267139 & -154887 \\
        Baseline3 (Greedy-Delay)  & 13565 & 4052 \\
        Baseline4 (TransformerPPO) & -4050226 & -3928943 \\
        Baseline5 (DiffusionRL)  & 32077 & 25511 \\ \bottomrule
    \end{tabular}
\end{minipage}
\hfill
\begin{minipage}{0.48\textwidth}
    \caption{\small Performance under different number of edge servers}
    \label{tab:raw_rewards_N_configs_}
    \centering
    \begin{tabular}{@{}lrrr@{}}
        \toprule
        \textbf{Algorithm}  & \textbf{N=15} & \textbf{N=20} \\ \midrule
        Ours  & \textbf{21080} & \textbf{28961} \\
        Baseline1 (Greedy-Accuracy) & -219696 & -278269 \\
        Baseline2 (Greedy-Compute)  & -181459 & -125089 \\
        Baseline3 (Greedy-Delay)  & 7235 & 10928 \\
        Baseline4 (TransformerPPO)  & -5642526 & -76411 \\
        Baseline5 (DiffusionRL)  & 14823 & 17939 \\ \bottomrule
    \end{tabular}
\end{minipage}
\end{table*}

\subsection{Performance Bound}

The core of our approach is to minimize an upper bound of the drift-plus-penalty expression at each timeslot $t$. The algorithm is designed to make a decision $\vect{a}(t)$ at each slot to greedily minimize the right-hand side of the drift-plus-penalty inequality. Therefore, the resulting value is no larger than the value achieved by any other policy, including the optimal stationary policy $\vect{a}^*(t)$. Taking conditional expectation on $\vect{\Theta}(t)$ yields:
\begin{align}
&\mathbb{E}[V \zeta(t) + \Delta(\vect{\Theta}(t)) | \vect{\Theta}(t)] \nonumber \\
&\leq \mathbb{E}\left[B + V \zeta(t) + \sum_{j=1}^{N+U} Q_j(t) y_j(t) \Big| \vect{\Theta}(t)\right] \\
&\leq B + \mathbb{E}\left[V \zeta(\vect{\omega}(t), \vect{a}^*) + \sum_{j=1}^{N+U} Q_j(t) y_j(\vect{\omega}(t), \vect{a}^*) \Big| \vect{\Theta}(t)\right].
\end{align}
Since $\vect{a}^*$ is independent of the queue backlogs $\vect{\Theta}(t)$, we can separate the expectations:
\begin{align}
&\mathbb{E}[V \zeta(t) + \Delta(\vect{\Theta}(t)) | \vect{\Theta}(t)] \nonumber \\
&\leq B + V \mathbb{E}[\zeta(\vect{\omega}(t), \vect{a}^*)] + \sum_{j=1}^{N+U} Q_j(t) \mathbb{E}[y_j(\vect{\omega}(t), \vect{a}^*)] \\
&\leq B + V \zeta^*
\end{align}
where the final inequality holds because $\mathbb{E}[y_j(\vect{\omega}(t), \vect{a}^*)] \leq 0$ for all $j$.

Taking the expectation over $\vect{\Theta}(t)$ and summing over $t \in \{0, \dots, T-1\}$:
\begin{equation}
\sum_{t=0}^{T-1} \mathbb{E}[\Delta(\vect{\Theta}(t))] + V \sum_{t=0}^{T-1} \mathbb{E}[\zeta(t)] \leq (B + V\zeta^*)T .
\end{equation}
The sum of the drifts forms a telescoping series:
\begin{equation}
\sum_{t=0}^{T-1} \mathbb{E}[\Delta(\vect{\Theta}(t))] = \mathbb{E}[L(\vect{\Theta}(T))] - \mathbb{E}[L(\vect{\Theta}(0))].
\end{equation}
Assuming the system starts with empty virtual queues, $\vect{Q}(0) = \vect{0}$, which implies $L(\vect{\Theta}(0)) = 0$. Since the Lyapunov function $L(\vect{\Theta}(T)) \ge 0$, we have:
\begin{equation}
V \sum_{t=0}^{T-1} \mathbb{E}[\zeta(t)] \leq (B + V\zeta^*)T + \mathbb{E}[L(\vect{\Theta}(0))] = (B + V\zeta^*)T.
\end{equation}
Dividing by $VT$ and taking the limit as $T \to \infty$, we establish the performance bound of our proposed algorithm:
\begin{equation}
\lim_{T \to \infty} \frac{1}{T} \sum_{t=0}^{T-1} \mathbb{E}[\zeta(t)] \leq \zeta^* + \frac{B}{V}.
\end{equation}
This result demonstrates that the long-term time-averaged cost achieved by our algorithm is within a constant $B/V$ of the optimal cost $\zeta^*$. By choosing a sufficiently large control parameter $V$, our algorithm's performance can be made arbitrarily close to the optimum.

\subsection{Constraint Satisfaction and System Stability}

We now prove that the long-term computational constraints are satisfied. This is equivalent to showing that the virtual queues $Q_j(t)$ are mean rate stable. We make a standard assumption for such proofs, often referred to as Slater's condition: there exists a stationary policy $\vect{a}^{\epsilon}(t)$ and a constant $\epsilon > 0$ such that for all $j$:
\begin{equation}
\mathbb{E}[y_j(\vect{\omega}(t), \vect{a}^{\epsilon}(t))] \leq -\epsilon, \quad \forall j \in \{1, \dots, N+U\}.
\end{equation}
This implies that there is a feasible policy that keeps the system strictly within its capacity constraints. We also assume $\zeta(t)$ is bounded, i.e., $\zeta_{\min} \leq \zeta(t) \leq \zeta_{\max}$.

Following the same logic as in the performance bound analysis, but using the policy $\vect{a}^{\epsilon}$ instead:
\begin{align}
&\mathbb{E}[\Delta(\vect{\Theta}(t)) | \vect{\Theta}(t)] + V \mathbb{E}[\zeta(t) | \vect{\Theta}(t)] \nonumber \\ 
&\leq B + V \mathbb{E}[\zeta(\vect{\omega}(t), \vect{a}^{\epsilon})] + \sum_{j=1}^{N+U} Q_j(t) \mathbb{E}[y_j(\vect{\omega}(t), \vect{a}^{\epsilon})].
\end{align}
Using the bounds on $\zeta(\cdot)$ and the Slater condition:
\begin{align}
&\mathbb{E}[\Delta(\vect{\Theta}(t)) | \vect{\Theta}(t)] + V \zeta_{\min} \leq B + V \zeta_{\max} - \epsilon \sum_{j=1}^{N+U} Q_j(t) \\
&\mathbb{E}[L(\vect{\Theta}(t+1)) - L(\vect{\Theta}(t)) | \vect{\Theta}(t)] \nonumber \\ 
&\leq B + V(\zeta_{\max} - \zeta_{\min}) - \epsilon \sum_{j=1}^{N+U} Q_j(t).
\end{align}
Let $H \triangleq B + V(\zeta_{\max} - \zeta_{\min})$. Taking expectation over $\vect{\Theta}(t)$ and summing from $t=0$ to $T-1$:
\begin{align}
\mathbb{E}[L(\vect{\Theta}(T))] - \mathbb{E}[L(\vect{\Theta}(0))] \leq HT - \epsilon \sum_{t=0}^{T-1} \sum_{j=1}^{N+U} \mathbb{E}[Q_j(t)].
\end{align}
With $L(\vect{\Theta}(0)) = 0$ and $L(\vect{\Theta}(T)) \geq 0$, we can rearrange the terms:
\begin{equation}
\sum_{t=0}^{T-1} \sum_{j=1}^{N+U} \mathbb{E}[Q_j(t)] \leq \frac{H \cdot T}{\epsilon}.
\end{equation}
From the queue update rule $Q_j(t+1) = \max\{Q_j(t) + y_j(t), 0\}$, we have $Q_j(t+1) \ge Q_j(t) + y_j(t)$. Summing over $t$ yields $\sum_{t=0}^{T-1} y_j(t) \leq Q_j(T) - Q_j(0) = Q_j(T)$. Taking expectations and dividing by $T$:
\begin{equation}
\frac{1}{T} \sum_{t=0}^{T-1} \mathbb{E}[y_j(t)] \leq \frac{\mathbb{E}[Q_j(T)]}{T}. \label{eq:y_q_relation}
\end{equation}
From the drift inequality, $\mathbb{E}[L(\vect{\Theta}(T))] \leq HT$. By the Cauchy-Schwarz inequality:
\begin{align}
\left( \sum_{j=1}^{N+U} \mathbb{E}[Q_j(T)] \right)^2
    &\leq (N+U) \sum_{j=1}^{N+U} \mathbb{E}[Q_j(T)]^2 \notag \\
    &\leq (N+U) \sum_{j=1}^{N+U} \mathbb{E}[Q_j(T)^2].
\end{align}
This leads to:
\begin{equation}
\left( \sum_{j=1}^{N+U} \mathbb{E}[Q_j(T)] \right)^2 \leq 2(N+U) \mathbb{E}[L(\vect{\Theta}(T))] \leq 2(N+U)HT.
\end{equation}
Therefore,
\begin{equation}
\frac{1}{T} \sum_{j=1}^{N+U} \mathbb{E}[Q_j(T)] \leq \sqrt{\frac{2(N+U)H}{T}}.
\end{equation}
As $T \to \infty$, the right-hand side goes to 0. Since $\mathbb{E}[Q_j(T)] \ge 0$, this implies:
\begin{equation}
\lim_{T \to \infty} \frac{\mathbb{E}[Q_j(T)]}{T} = 0, \quad \forall j \in \{1, \dots, N+U\}.
\end{equation}
Substituting this result into \eqref{eq:y_q_relation} proves that the long-term computational constraints are satisfied:
\begin{equation}
\lim_{T \to \infty} \frac{1}{T} \sum_{t=0}^{T-1} \mathbb{E}[y_j(t)] \leq 0, \quad \forall j.
\end{equation}
This concludes the proof, showing that our algorithm not only achieves near-optimal performance but also guarantees system stability and constraint satisfaction.

\ifCLASSOPTIONcaptionsoff
  \newpage
\fi

\section{Experiment results}

\subsection{Experiment Setting}

Most previous works simply use mathematical distributions (e.g., Poisson processes) to simulate and generate LLM requests, which fail to reflect the complex patterns of real-world workloads. In this work, we use real LLM query traces from the Alibaba Bailian (an AI model service platform)\cite{xiang2025servegen}. We use a subset of the overall customer request information, selected from a specific time period. 

We consider a heterogeneous edge-cloud architecture: edge servers and cloud servers, where $K=3$ distinct types of tasks are offloaded and processed over $T=100$ time frames. The computational capabilities of edge servers and cloud servers are initialized to reflect practical heterogeneity, with edge servers compute resources $f_n$ randomly sampled within $[2.5, 5]$ and cloud servers' $f_u$ within $[5, 7.5]$.  Task characteristics are parameterized by delay sensitivity $\alpha$ and accuracy sensitivity $\beta$, both sampled uniformly in $[0.5, 1.0]$, capturing diverse QoS requirements. The computational requirements for LLM inference are evaluated at two different model scales. For the small-scale model, prefill and decoding stages require 2 and 1 computation units, respectively; for the large-scale model, 8 and 4 computation units are needed for the same stages. The network latency is structured so that edge servers exhibit lower communication delays while cloud servers have higher delays, aligning with realistic network dynamics. Task accuracy exhibits a similar dichotomy: edge servers offer lower accuracy $[0.1, 0.5]$ compared to cloud servers $[0.6, 1.0]$, thus capturing quality-performance trade-offs across tiers.

In this section, we conduct extensive experiments to verify the effectiveness of our algorithm. 
In the token prediction part, we compare of our algorithm with the following baselines.
\begin{itemize}
    \item \texttt{LoRA}: Most widely used method to conduct efficient tuning \cite{hu2022lora}. We use ModernBert as the pretrained model and conduct fine-tuning.
    \item \texttt{LSTM}: Long Short-Term Memory (LSTM) is a type of recurrent neural network architecture designed to effectively process sequential data.\cite{graves2012long}
    \item \texttt{Transformers}: Transformers are a type of neural network architecture that uses self-attention mechanisms to process and model sequential data.\cite{vaswani2017attention}
    \item \texttt{Qwen2.5-7B}: This is a powerful large language model  created by Alibaba.\cite{qwen2025qwen25technicalreport}
    
\end{itemize}
In the task offloading part, we compare of our algorithm with the following baselines.

\begin{itemize}
    \item \texttt{Greedy\_Accuracy}: Offloading the queries to the device that can provide the highest accuracy.
    \item \texttt{Greedy\_Compute}: Offloading the queries to the device that has the highest computation power.
    \item \texttt{Greedy\_Delay}: Offloading the queries to the device that has the lowest delay.
    \item \texttt{DiffusionRL}: Using the Diffusion RL to conduct network optimization. This idea is widely used in \cite{du2024enhancing},\cite{du2024diffusion} \cite{du2023ai}. We additionally add the Lyapunov optimization to satisfy long-term constraints.
    \item \texttt{TransformerPPO}: Using the Transformer\cite{vaswani2017attention} as the neural network architecture and PPO\cite{schulman2017proximal} as the RL algorithm. We additionally add the Lyapunov optimization to satisfy long-term constraints.
\end{itemize}

\subsection{Experiment Results}
\textbf{Performance comparison of LOO and other algorithms.}
Tab. \ref{tab:raw_rewards_u_configs} and Tab. \ref{tab:raw_rewards_N_configs_} illustrate the performance of our algorithm and various baselines under different configurations of cloud servers (U) and edge servers (N). Specifically, we evaluate performance with U set to 15 and 20, and N set to 15 and 20. We use the Lyapunov reward as the metric, which consists of the QoE of clients and the queue that represents the satisfaction of the long-term constraints.
The results demonstrate that LOO consistently achieves the highest performance across all evaluated configurations. Specifically, LOO achieves rewards of 36,602 in (N = 4,U = 15), 30,757 in (N = 4,U = 20), 21,080 in (N = 15, U = 6) and 28,961 in (N = 20, U = 6). This performance significantly outperforms not only the Greedy-Accuracy, Greedy-Compute, and Greedy-Delay algorithms, but also exceeds reinforcement learning algorithms that require substantial training overhead. 

\begin{figure}[t]
    \centering 
    
    \begin{subfigure}[b]{0.24\textwidth}
        \centering
        \includegraphics[width=\textwidth]{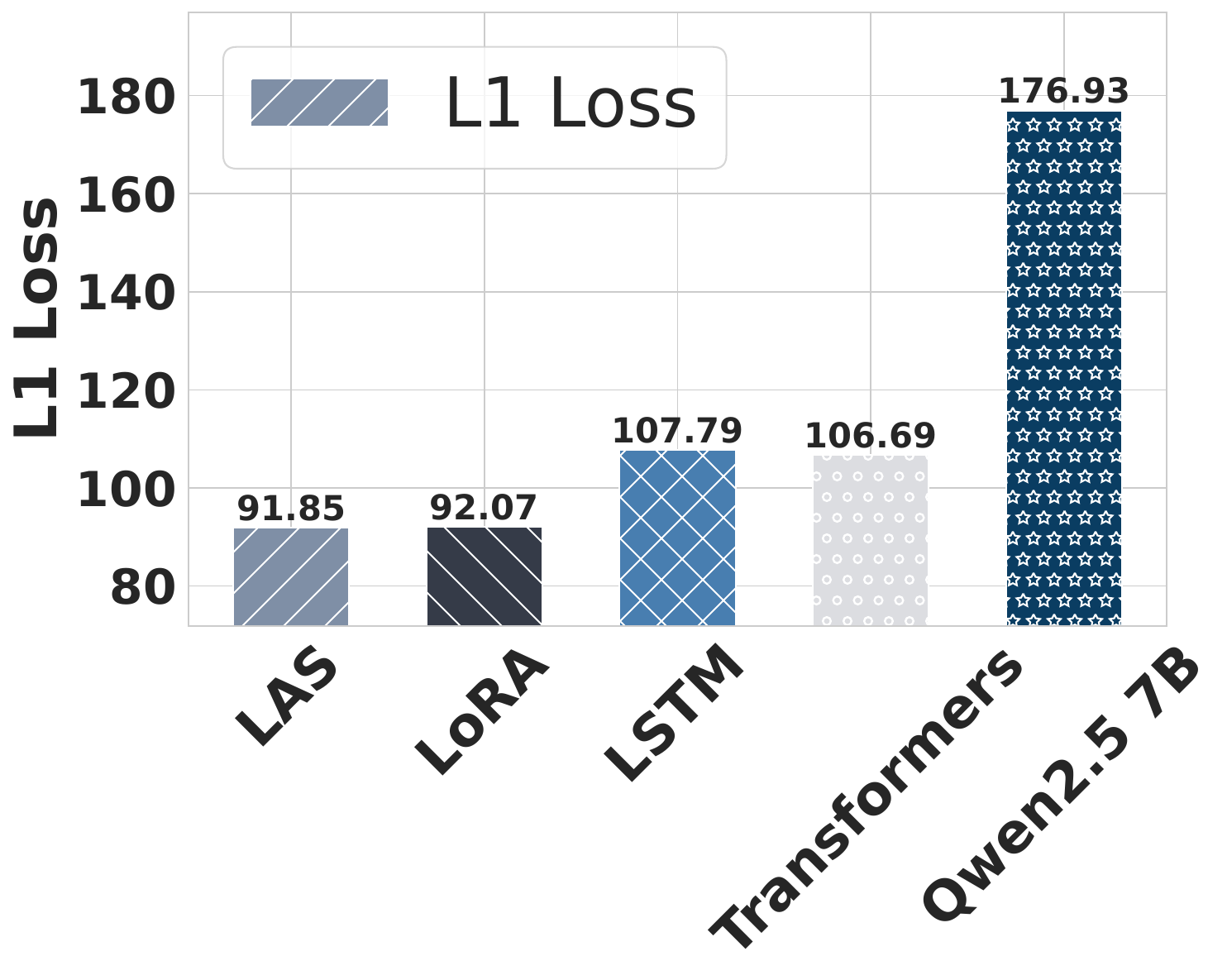}
        \caption{Performance comparison}
        \label{fig:performance_comp}
    \end{subfigure}
    \hfill 
    \begin{subfigure}[b]{0.24\textwidth}
        \centering
        \includegraphics[width=\textwidth]{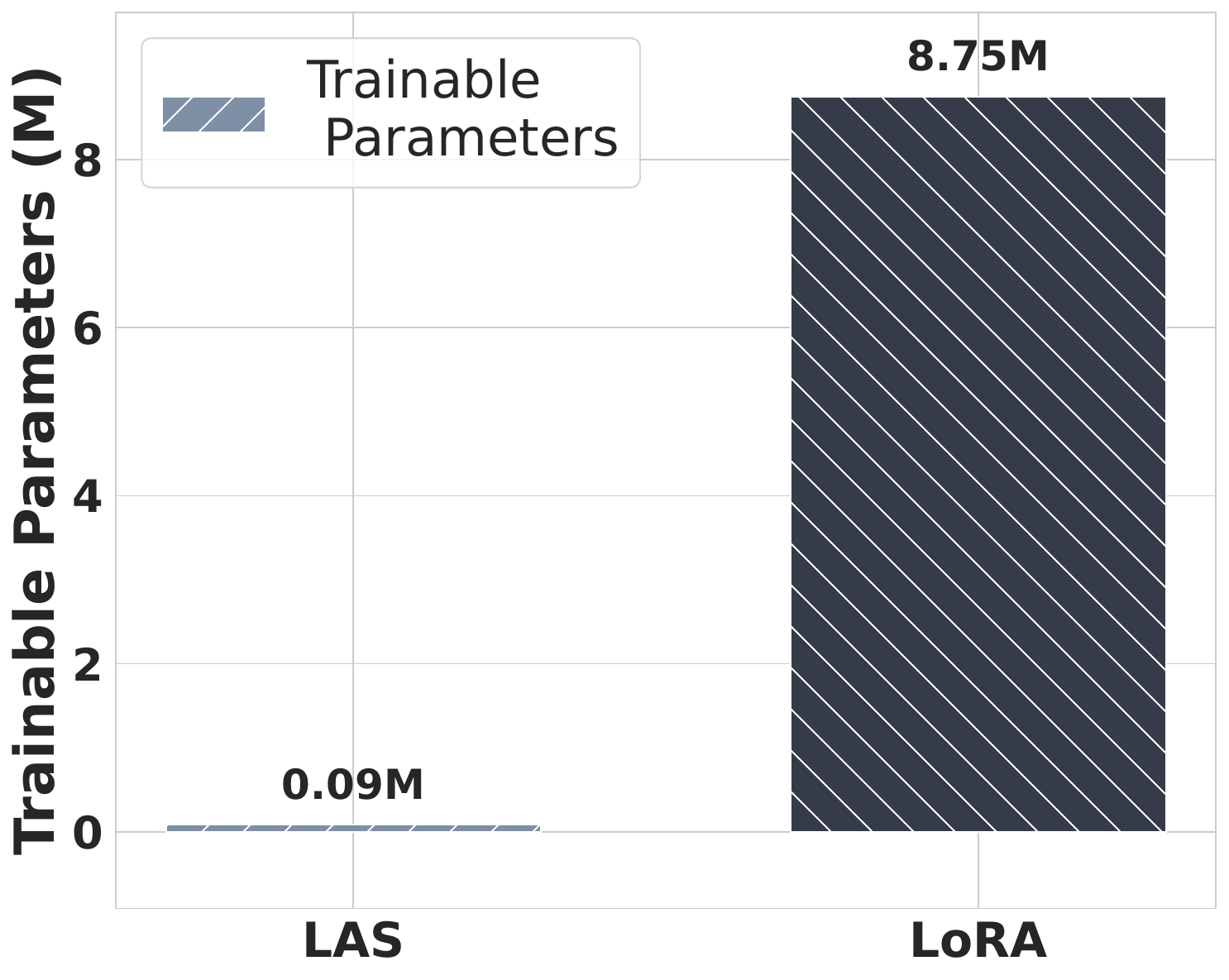}
        \caption{Trainable parameters}
        \label{fig:parameter_num_comp}
    \end{subfigure}

    \caption{Performance (L1 loss) and trainable parameter comparison of LAS and other baselines}
    \label{fig:enhanced_comparison}
\end{figure}

\textbf{Performance comparison of different token prediction method.}
Fig. \ref{fig:performance_comp} illustrates the comparison of the token prediction precision of our algorithm and other basellines. From the result, we can find that the l1 loss of the LAS is 91.85 while the L1 loss of other baselines is 92.07, 107.79, 106.69, 176.93, which is $0.2\%, 17.3\%, 16.15\%, 92.62\%$ higher. ModernBERT utilizes advanced rotary position embeddings (RoPE)\cite{su2024roformer} for enhanced position awareness and has been trained on trillions of diverse tokens, enabling broad and representative natural language understanding. The LAS module we design can dynamically re-weighting semantic features with the LAS module. The model highlights cues most relevant to output length while suppressing irrelevant information, leading to more focused predictions. This selective recalibration enhances the model’s sensitivity to length-related instructions in the prompt, significantly improving token count prediction accuracy. While the LoRA \cite{hu2022lora} algorithm fail to achieves this. As depicted in Fig. \ref{fig:parameter_num_comp}, the trainable parameter of the LoRA algorithm is 8.75M, while in LAS, we only introduce 0.09M activated parameters. which is $99\%$ lower while achieving better results. For the LSTM and transformer, because of lacking extensive pretrained semantic knowledge, these methods cannot achieve precise token prediction. For the Qwen2.5-7B model, although with pretrained knowledge, it cannot fit the need of the token prediction task.

\begin{table}[t]
\centering
\begin{minipage}{0.48\textwidth}
    \caption{Ablation study for token length predictor}
    \label{tab:ablation_study}
    \centering
    \begin{tabular}{@{}lrrr@{}}
        \toprule
        \textbf{Configuration} & \textbf{With predictor  } & \textbf{Without predictor  } \\ \midrule
        N=4,U=6 & 10808 & 5409  \\
        N=4,U=8 & 14324  & 11519  \\
        N=4,U=10 & 18339 & 14457  \\
        \bottomrule
    \end{tabular}
\end{minipage}
\end{table}

\textbf{Impact of token length predictor.}
We evaluated our algorithm under configurations (N=4, U=6), (N=4, U=8), and (N=4, U=10) with and without the token length predictor module. As shown in Table \ref{tab:ablation_study}, integrating the predictor substantially enhanced evaluation rewards: at (N=4, U=6), reward increased by 99.8\% from 5,409 to 10,808; at (N=4, U=8), it rose by 24.3\% from 11,519 to 14,324; and at (N=4, U=10), it improved by 26.9\% from 14,457 to 18,339. This significant improvement occurs because the output token length substantially impacts both the computational latency and workload of requests across the devices. Accurate prediction enables the algorithm to make more optimal offloading decisions.

\section{Conclusion}
In this paper, we introduced Argus, a novel framework to address the unpredictable inference latency of LLMs in dynamic, heterogeneous edge-cloud systems. This variability, caused by the autoregressive generation process, is a critical challenge that most existing resource allocation strategies overlook.

Our solution features two core components. First, the Length-Aware Semantics (LAS) module accurately predicts output token lengths before generation, enabling precise workload profiling. Building on this, our Lyapunov-guided Offloading Optimization (LOO) module formulates task allocation as a long-term Quality-of-Experience maximization problem. This is efficiently solved by our low complexity novel Iterative Offloading Algorithm with Damping and Congestion Control (IODCC). Finally, We provided a rigorous theoretical analysis proving Argus achieves near-optimal performance while guaranteeing system stability. Our extensive experiments on real-world query traces further demonstrated that Argus significantly outperforms state-of-the-art baselines in various settings.

\bibliography{reference.bib}
\bibliographystyle{IEEEtran}
\end{document}